\documentstyle[aps,prl,preprint,floats,epsfig]{revtex}

\begin{document}

\preprint{\tighten\vbox{\hbox{\hfil CLNS 99/1658}
                        \hbox{\hfil CLEO 99-22}
}}

\title{ Measurement of the {\boldmath $ B^0$}  and  {\boldmath $ B^+$} meson  masses from {\boldmath $B^0 \to \psi^{(\prime)} \, K^0_S$}  and  {\boldmath $B^+ \to \psi^{(\prime)} \, K^+$}  decays}

\author{CLEO Collaboration}
\date{\today}

\maketitle
\tighten

\begin{abstract} 
Using a  sample of   $9.6\times10^6$ $B \overline B$ meson pairs collected with the CLEO detector,  we have fully reconstructed 135 $B^0 \to \psi^{(\prime)}  \, K^0_S$ and 526 $B^+ \to \psi^{(\prime)} \, K^+$ candidates with very low background. 
 We fitted the $\psi^{(\prime)} \, K$ invariant mass distributions of these $B$ meson candidates and  measured  the masses of the neutral and charged  $B$ mesons to be  $M(B^0)=5279.1 \pm 0.7[{\rm stat}] \pm 0.3[{\rm syst}]$~MeV/$c^2$ and $M(B^+)=5279.1 \pm 0.4[{\rm stat}]  \pm 0.4[{\rm syst}]$~MeV/$c^2$. The precision is a  significant improvement over previous measurements. 

\end{abstract}
\newpage

{
\renewcommand{\thefootnote}{\fnsymbol{footnote}}

\begin{center}
S.~E.~Csorna,$^{1}$ I.~Danko,$^{1}$ K.~W.~McLean,$^{1}$
Sz.~M\'arka,$^{1}$ Z.~Xu,$^{1}$
R.~Godang,$^{2}$ K.~Kinoshita,$^{2,}$%
\footnote{Permanent address: University of Cincinnati, Cincinnati OH 45221}
I.~C.~Lai,$^{2}$ S.~Schrenk,$^{2}$
G.~Bonvicini,$^{3}$
% ...
\end{center}
 
\small
\begin{center}
$^{1}${Vanderbilt University, Nashville, Tennessee 37235}\\
$^{2}${Virginia Polytechnic Institute and State University,
Blacksburg, Virginia 24061}\\
$^{3}${Wayne State University, Detroit, Michigan 48202}\\
% ...
\end{center}

\setcounter{footnote}{0}
}
\newpage

\newcounter{syst_counter}
 The previous measurements of the $B$ meson masses at $e^+e^-$ colliders operating
at $\Upsilon(4S)$ energy~\cite{B-mass-ARGUS-CLEO}  were obtained from fits to  the distributions of the 
beam-constrained $B$ mass, defined as $M_{\rm bc} \equiv \sqrt{E^2_{\rm beam}-p^2(B)}$, where $p(B)$ is the absolute value of the $B$ candidate momentum. 
Substitution of the beam energy  for the measured  energy of the $B$ meson candidate   results 
in a significant improvement of the mass resolution, therefore 
the beam-constrained mass  method is  the  technique of choice  
for the $M(B^0)-M(B^+)$ mass difference measurement. 
However, the precision of the measurement of the absolute $B^0$ and $B^+$ meson masses is limited by the systematic uncertainties in the absolute beam energy scale and in the  correction for initial state radiation.  
For this measurement  we selected $B^0 \to \psi^{(\prime)} \, K^0_S$ and  
$B^+ \to \psi^{(\prime)} \, K^+$  candidates~\cite{charge-conjugate}, reconstructing $\psi^{(\prime)} \to \ell^+ \ell^-$ and $K^0_S \to \pi^+ \pi^-$ decays. We used  both $e^+e^-$ and $\mu^+\mu^-$ modes for the $\psi^{(\prime)}$ reconstruction.  We then determined  the $B^0$ and $B^+$ meson masses by fitting  the $\psi^{(\prime)} \, K^0_S$ and $\psi^{(\prime)} \, K^+$ invariant mass distributions.
The main reasons to use  the  $\psi^{(\prime)} K$ rather than more copious  $D^{(*)} n\pi$ final states are, first, that the background is very low; second, 
that  the $J/\psi$ and $\psi(2S)$ mesons are  heavy, and their masses are  very well measured~\cite{PDG}.
As discussed below, constraining the reconstructed $J/\psi$ and $\psi(2S)$  
masses to their world average values makes our $B$ mass measurement 
insensitive to imperfections in the lepton momentum reconstruction. 
By comparing the beam-constrained $B$ mass to the  
$B^0$ and $B^+$  mass values obtained in our measurement, one  
 could  set the absolute  beam energy scale  at the $e^+e^-$ 
colliders operating in  the $\Upsilon(4S)$ energy region. 

The data were collected at   the Cornell Electron Storage Ring  (CESR)
with    two   configurations    of     the   CLEO detector,     called
CLEO~II~\cite{Kubota:1992ww}  and  CLEO~II.V~\cite{Hill:1998ea}.
The components of the CLEO detector most relevant to this analysis are the charged particle tracking system, the CsI electromagnetic calorimeter, and the muon chambers.
In CLEO~II, the momenta of charged particles are measured in a tracking system consisting of a 6-layer straw tube chamber,  10-layer precision drift chamber, and 51-layer main drift chamber, all operating inside a 1.5 T  solenoidal magnet. The main drift chamber also provides a measurement of the  specific ionization, $dE/dx$, used for particle identification.  For  CLEO~II.V, the straw tube  chamber was replaced  with a  3-layer silicon vertex detector. The muon chambers  consist of proportional counters placed at various depths in the steel absorber.
Track fitting was performed using a Kalman filtering technique, first applied to track fitting by P.~Billoir~\cite{Billoir:1984mz}. The track fit   sequentially adds  the measurements provided by the tracking system   to correctly  take
into account multiple scattering and energy loss of a particle in the
detector material. For each physical track, separate fits are   performed  using different particle hypotheses.

 For this measurement we used   9.1~$\rm fb^{-1}$ of $e^+e^-$ data taken at the $\Upsilon(4S)$ energy and 4.4~$\rm fb^{-1}$ recorded 60~MeV below the  $\Upsilon(4S)$ energy. Two thirds of the data used were collected with the CLEO~II.V detector.
 All of the simulated  event samples used in this analysis were
generated with a GEANT-based~\cite{GEANT} simulation of the 
CLEO detector response and  were processed in a similar manner as the  data.  

 Electron candidates were identified based on the ratio of the track momentum to the associated shower energy in the CsI calorimeter and specific ionization  in the drift chamber. We  recovered some of the  bremsstrahlung photons by selecting the photon  shower with the smallest opening angle with respect to the  direction of the $e^\pm$ track evaluated at the interaction point, and then   requiring this opening angle to be smaller than $5^\circ$. 
 For the $\psi^{(\prime)} \to \mu^+ \mu^-$ reconstruction one of the muon candidates was  required to penetrate the steel absorber to a depth greater than 3  nuclear interaction lengths.
We relaxed the absorber penetration requirement for  the second muon candidate  if it  pointed to the end-cap muon chambers and its energy  was too low to reach a counter.  For these muon candidates  we required the ionization  signature in the  CsI calorimeter to be  consistent with that of a muon.
 
\begin{figure}[htbp]
\centering
\epsfxsize=16cm
\epsfbox{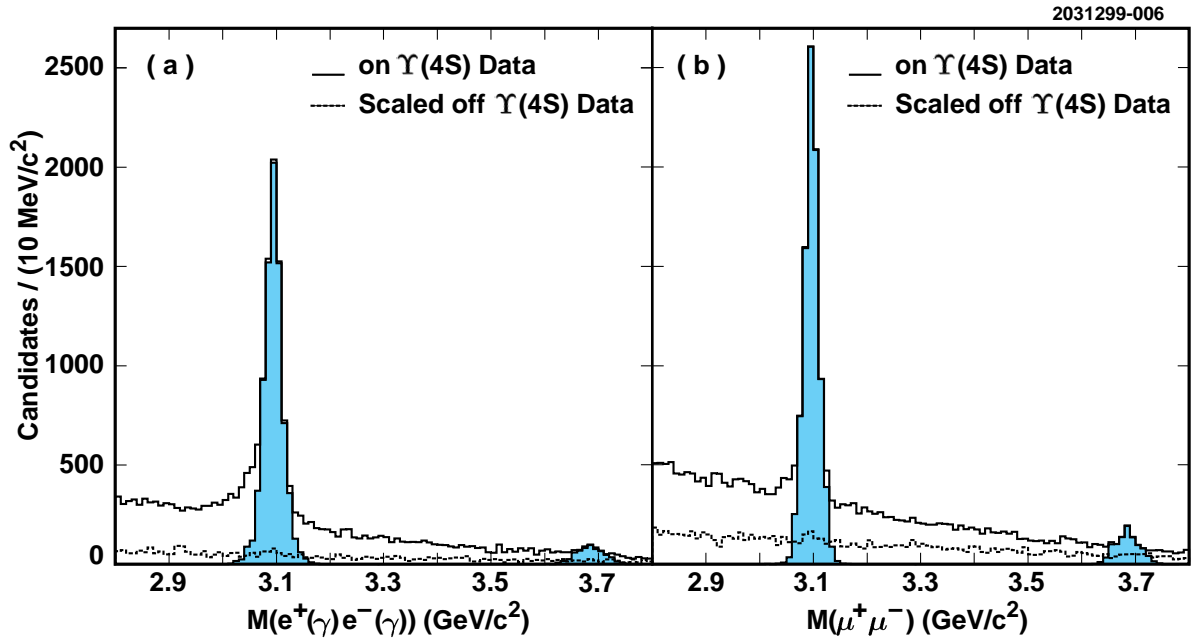}
\caption{(a) $\psi^{(\prime)} \to e^+ e^-$ and  (b) $\psi^{(\prime)} \to \mu^+ \mu^-$ 
 candidates  in data. The solid line represents the data taken at  $\Upsilon(4S)$ energy; the dashed line represents the scaled  off resonance data  showing the level of background from non-$B \overline B$ events. The shaded parts of the histograms  represent the $\psi^{(\prime)}$ candidates 
with the absolute value of the normalized invariant mass less than 3.}
\label{fig:psi_psiprime_data}
\end{figure}

We extensively used normalized variables, taking  advantage of well-understood track and photon-shower four-momentum covariance matrices to calculate the expected resolution for each combination. The use of normalized variables allows uniform candidate selection criteria to be applied to the data collected with the CLEO~II and CLEO~II.V detector configurations.
The  $e^+(\gamma) e^-(\gamma)$ and $\mu^+ \mu^-$  invariant mass distributions for the $\psi^{(\prime)} \to \ell^+ \ell^-$ candidates in data are  shown in Fig.~\ref{fig:psi_psiprime_data}.
We selected the  $\psi^{(\prime)} \to \ell^+ \ell^-$ signal candidates requiring the absolute value of the normalized invariant mass to be less than $3$.
The average $\ell^+ \ell^-$ invariant mass resolution  is approximately   12~MeV$/c^2$.
For each $\psi^{(\prime)}$ candidate  we  performed a fit constraining its  mass to the world average  value~\cite{PDG}. This mass-constraint fit improves 
the $\psi^{(\prime)}$ energy resolution  almost by a factor of 4 and  the  absolute  momentum resolution by 30\%. 
\

The $K^0_S$ candidates were selected from  pairs of tracks forming 
well-measured displaced vertices.  The daughter pion tracks were re-fitted taking into account the position of the displaced vertex and  were constrained to originate from the same spatial point.
 The resolution in $\pi^+ \pi^-$ invariant mass is approximately 4~MeV$/c^2$. 
After requiring the absolute value of the normalized $\pi^+ \pi^-$ invariant mass to be less than  3,  we  performed a fit constraining the mass of each $K^0_S$ candidate to the world average  value~\cite{PDG}.

\begin{figure}[htbp]
\centering
\epsfxsize=16cm
\epsfbox{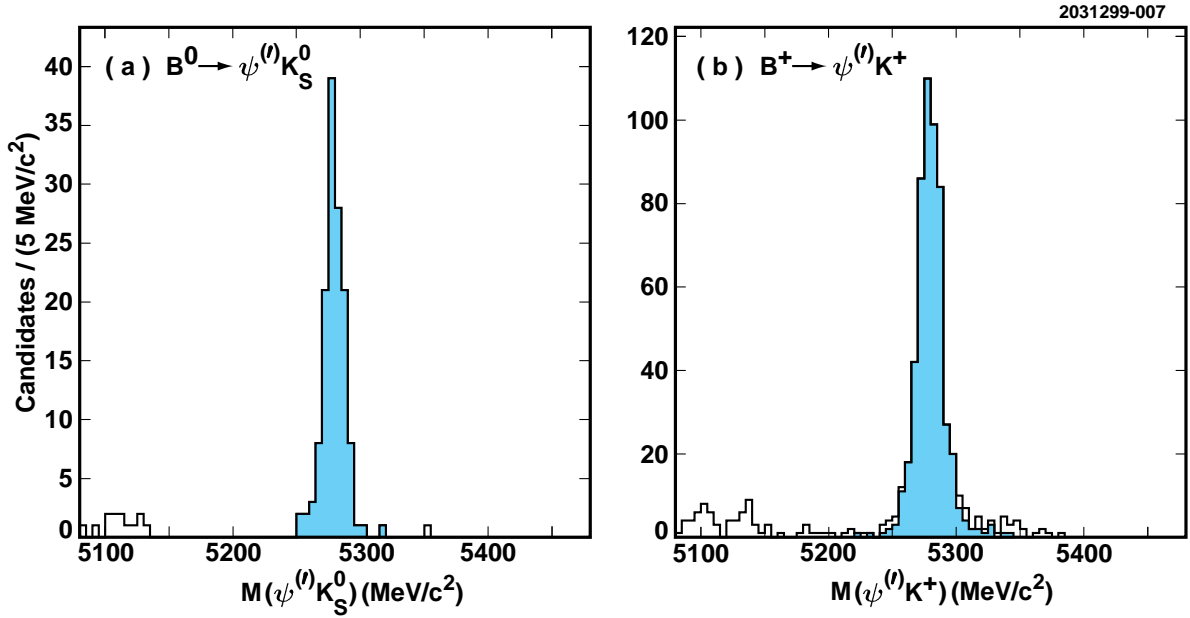}
\caption{The invariant mass distributions for (a) $B^0 \to \psi^{(\prime)}\, K^0_S$  and   (b) $B^+ \to \psi^{(\prime)}\, K^+$  candidates passing 
the beam-constrained $B$ mass requirement. The shaded parts of the histograms represent    the candidates selected for the $B$ mass measurement fits by requiring that the absolute value of the normalized invariant mass to be less than 4 in (a) and 3 in (b).}      
\label{fig:invm_data}
\end{figure}

 The   $B \to  \psi^{(\prime)} \, K$ candidates were selected by means of two observables.
The first observable is the beam-constrained $B$ mass 
$M_{\rm bc} \equiv \sqrt{E^2_{\rm beam}-p^2(B)}$. The resolution in $M_{\rm bc}$ for  the $B \to  \psi^{(\prime)} \, K$ candidates is approximately  2.7~MeV/$c^2$ and   is dominated by the beam energy spread.  We required $|M_{\rm bc}-5280\;{\rm MeV}/c^2|/\sigma(M_{\rm bc})<3$. The requirement on $M_{\rm bc}$, equivalent to a requirement on the absolute value of the $B$ candidate momentum, is  used only for background suppression, and  does  not bias the $B$ mass measurement.
 The second observable is the invariant mass of the $\psi^{(\prime)} \, K$ 
system. The average resolutions in $M(\psi^{(\prime)} \,  K^0_S)$ and $M(\psi^{(\prime)} \, K^+)$ are respectively  8~MeV/$c^2$ and  11~MeV/$c^2$. 
The $M(\psi^{(\prime)} \, K)$ distributions for the candidates passing the beam-constrained $B$ mass requirement are shown in Fig.~\ref{fig:invm_data}. To select signal candidates, we required 
$|M(\psi^{(\prime)} \,  K^0_S)-5280\;{\rm MeV}/c^2|/\sigma(M)<4$ and $|M(\psi^{(\prime)} \, K^+)-5280\;{\rm MeV}/c^2|/\sigma(M)<3$; the allowed invariant mass intervals are sufficiently wide  not to introduce bias in the $B$ mass measurement. 
 These selections yielded 135 $B^0 \to \psi^{(\prime)}\, K^0_S$ candidates: 125 in the $B^0 \to J/\psi\, K^0_S$ mode  and 10 in the $B^0 \to \psi(2S)\, K^0_S$ mode. We  estimated the background to be  $0.13^{+0.09}_{-0.05}$ events. The 
selections yielded  526 $B^+ \to \psi^{(\prime)}\, K^+$ candidates: 468 in $B^+ \to J/\psi\, K^+$ mode  and 58 in $B^+ \to \psi(2S) \,K^+$ mode.  The background from $B^+ \to \psi^{(\prime)} \, \pi^+$ decays was estimated to be $0.9\pm0.3$  events, whereas  all other background sources were estimated to contribute  $2.3^{+1.0}_{-0.5}$ events. The backgrounds were evaluated  with simulated events and the data recorded at the energy below the  $B \overline B$ production threshold. We discuss the  systematics associated with background below, together with other systematic uncertainties.  
\

The $B$ meson masses were extracted from the $\psi^{(\prime)}\, K^0_S$ and 
 $\psi^{(\prime)}\, K^+$ invariant mass distributions with an unbinned likelihood fit. The likelihood function is 
\begin{eqnarray}\label{eq:likelihood}
\nonumber
{\cal  L}(M(B), S) = \prod_i  
\frac{G(M_i-M(B)|S\sigma_i)}
{\int G(M-M(B)|S\sigma_i) dM } \; \; \;,    
\end{eqnarray}
where $M_i$ is the invariant mass of a $\psi^{(\prime)}\, K$ combination, $\sigma_i$ is the calculated invariant mass uncertainty for that $\psi^{(\prime)}\, K$ combination, and $G(x|\sigma)\equiv 1/(\sqrt{2\pi}\sigma)\exp(-x^2/2\sigma^2)$.
The product is over the $B^0$  or $B^+$ meson candidates. The parameters of the fit are the $B$ meson mass $M(B)$ and  a global scale factor $S$ that modifies the calculated invariant-mass uncertainties  $\sigma_i$.
The integration limits of  the normalization integral in the denominator correspond to the  signal regions  defined for  the 
$\psi^{(\prime)}\, K$ invariant mass distributions: [5280~MeV/$c^2 \pm 4\sigma(M)$] for $B^0$ and   [5280~MeV/$c^2\pm 3\sigma(M)$] for $B^+$ candidates.
 From the fits to the $\psi^{(\prime)}\, K^0_S$ invariant-mass distribution,  we obtained $M(B^0)=5278.97 \pm 0.67$~MeV/$c^2$,  $S = 1.24 \pm 0.08$, and the 
correlation coefficient $\rho(M(B^0),S)=-0.013$. 
For $\psi^{(\prime)}\, K^+$ we obtained $M(B^+) = 5279.50 \pm 0.41$~MeV/$c^2$, $S = 1.09 \pm 0.04$, and $\rho(M(B^+),S)=-0.015$.
The  values of  the scale factor $S$ and uncertainty in $M(B)$  returned by the fits  are 
in good agreement with the values obtained from  simulated events.

Table~I   lists the  bias corrections together with associated 
systematic uncertainties, which we will discuss below. 

\begin{table}[htbp]
\label{tab:syst}
\caption{Bias corrections and associated systematic uncertainties. The total  systematic uncertainty was obtained by  adding in quadrature  all the  uncertainties listed in the table.}
\begin{center}
\begin{tabular}{l l r r} 
\multicolumn{2}{c}{Source of bias }  
 & \multicolumn{2}{c}{Correction and uncertainty} \\
& & \multicolumn{2}{c}{  (MeV/$c^2$) } \\
                                 &   &  $M(B^0)$ & $M(B^+)$ \\ \hline
\setcounter{syst_counter}{1}    
(\roman{syst_counter})  & Bias observed for simulated events  &  $+0.08\pm0.08$     & $-0.13\pm0.13$     \\
\stepcounter{syst_counter}    
(\roman{syst_counter})  & Background                          &  $\pm 0.05$     & $\pm 0.17$         \\
\stepcounter{syst_counter}
(\roman{syst_counter})  & $B$ mass likelihood fit             &  $\pm 0.14$     & $\pm 0.09$          \\
\stepcounter{syst_counter}
(\roman{syst_counter}) & $\psi^{(\prime)}$ four-momentum measurement   
&  $\pm 0.05$     & $\pm  0.06$ \\  
\stepcounter{syst_counter} 
(\roman{syst_counter})  &
$K^0_S$ four-momentum measurement    &  $\pm 0.28$     &  ---  \\ 
\stepcounter{syst_counter}    
(\roman{syst_counter}) & $K^+$ momentum measurement 
&  ---     &  $-0.32 \pm 0.34$  \\ 
\stepcounter{syst_counter}
(\roman{syst_counter})  & Detector misalignment              &  negligible     &  negligible   \\ \hline 
\multicolumn{2}{c}{Total correction and systematic uncertainty} & $+0.08 \pm 0.33$ & $ -0.45 \pm 0.42$ \\
\end{tabular}
\end{center} 
\end{table}

\setcounter{syst_counter}{1}    
(\roman{syst_counter}) {\it Measuring $B$ masses using  simulated events.} ---  Applying the same procedure as in  the data analysis, we 
measured the $B^0$ and $B^+$ mass using  30286 $B^0 \to \psi^{(\prime)}\, K^0_S$  and  34519 $B^+ \to \psi^{(\prime)}\, K^+$ candidates reconstructed from  a sample of simulated events. We obtained  $M(B^0)-M^{\rm input}_{B^0} = - 0.08 \pm 0.04$~MeV/$c^2$ and  $M(B^+)-M^{\rm input}_{B^+} = + 0.13 \pm 0.05$~MeV/$c^2$. We applied $+0.08$~MeV/$c^2$ and $-0.13$~MeV/$c^2$ corrections to the $B^0$ and $B^+$ mass values and assigned 100\% of those corrections  
as systematic uncertainties.

\stepcounter{syst_counter}    
(\roman{syst_counter}) {\it Background.} --- The estimated  mean background for the $B^0$ candidates is   $0.13^{+0.09}_{-0.05}$ events, we therefore conservatively  assumed the probability of finding  a background event  
in our  sample to be 22\%. The $B^0$ background  candidates are expected to be   uniformly distributed across the $B^0$ mass signal region. 
We performed the $B^0$ mass fits excluding the one  candidate  with the 
highest or the lowest 
normalized  $\psi^{(\prime)}\, K^0_S$ invariant  mass; the largest observed  $B^0$ mass shift was 0.25 MeV/$c^2$.  We   multiplied  this  0.25 MeV/$c^2$ shift by the 22\% probability of having a background event in our sample and 
assigned 0.05 MeV/$c^2$ as the systematic uncertainty in $B^0$ mass due to
background.
 For the $B^+$ signal, the background from $B^+ \to \psi^{(\prime)} \, \pi^+$ decays was estimated to be $0.9\pm0.3$  events, all other background sources were estimated to contribute  $2.3^{+1.0}_{-0.5}$ events. 
The $B^+$ background  candidates, with the exception of   $B^+ \to \psi^{(\prime)} \,\pi^+$ events, are expected to be   uniformly distributed across the $B$ mass signal region. 
The $B^+ \to \psi^{(\prime)} \,\pi^+$ events reconstructed as $B^+ \to \psi^{(\prime)} \,K^+$  produce high $\psi^{(\prime)} \,K^+$ 
invariant mass. 
  We performed the $B^+$ mass fits excluding 4 candidates  with the highest or the lowest normalized  $\psi^{(\prime)}\, K^+$ invariant  mass and assigned the largest shift of the measured $B^+$ mass (0.17 MeV/$c^2$) as the  systematic uncertainty in $B^+$ mass due to background.

\stepcounter{syst_counter}    
(\roman{syst_counter}) {\it $B$ mass likelihood fit.} --- We studied the systematics associated with  the unbinned likelihood fit procedure by changing the fit function from a Gaussian to sum of two Gaussians. We  also allowed the fit to determine different scale factors $S$ for 
the candidates coming from  CLEO~II and CLEO~II.V data, 
or for the candidates with  $\psi^{(\prime)} \to e^+ e^-$  and 
$\psi^{(\prime)} \to \mu^+ \mu^-$. We assigned the largest shift of the measured $B$ mass  as the  systematic uncertainty.

\stepcounter{syst_counter}    
(\roman{syst_counter}) {\it $\psi^{(\prime)}$ four-momentum measurement.} --- Even if the $B$ mass measurements using the simulated events show negligible bias, a  bias in the  measurement is still in principle  
possible  because of the 
uncertainty in the absolute  magnetic field scale, an imperfect description of the  detector material used by the Billoir fitter, or detector misalignment. 
 For the $\psi^{(\prime)}$ four-momentum measurement, these systematic effects   along  with the systematics associated with bremsstrahlung  are rendered negligible by the the $\psi^{(\prime)}$ mass-constraint fit. The measured  position of the $J/\psi$ mass peak  allows a reliable evaluation of the possible bias in the lepton momentum measurement.
We measured the positions of the $J/\psi \to \mu^+ \mu^-$ and $J/\psi \to e^+ e^-$ peaks by fitting the inclusive $\mu^+ \mu^-$ and $e^+(\gamma) e^-(\gamma)$ invariant mass distributions. In these fits  we used  the signal shapes derived from a high-statistics sample of simulated $J/\psi \to \ell^+ \ell^-$   events generated with the $J/\psi$ mass of 3096.88~MeV/$c^2$~\cite{PDG}.
In simulated  $J/\psi \to \mu^+ \mu^-$ events,  the reconstruction procedure introduces a bias of less  than 0.03~MeV/$c^2$  in the measured  $J/\psi$  mass. 
 We found that  the $J/\psi \to \mu^+ \mu^-$ peak was shifted 
by $+0.5\pm0.2$~MeV/$c^2$ in data compared to simulated events; the corresponding value of the  $J/\psi \to e^+ e^-$ peak shift was $+0.7\pm0.2$~MeV/$c^2$.
A $+0.5$~MeV/$c^2$ shift corresponds to overestimation of the lepton absolute momenta by  approximately 0.02\%. 
 A variation  of the lepton absolute momenta by 0.1\% produced a shift of less  than 0.02~MeV/$c^2$  in the measured $B$ mass. We therefore neglected the 
systematic uncertainty associated with the lepton momentum measurement. In addition, we varied the world average $J/\psi$ and  $\psi(2S)$ mass values used in 
the mass-constraint fits  by one standard deviation~\cite{PDG}; the resulting 0.05~MeV/$c^2$  and  0.06~MeV/$c^2$ shifts of the measured $B^0$ and  $B^+$ masses were assigned as systematic uncertainties.

\stepcounter{syst_counter} 
(\roman{syst_counter})  {\it $K^0_S$ four-momentum measurement} ---
The  systematic uncertainty of our $B$ mass measurement  is dominated by a possible bias  in the kaon  four-momentum measurement.
 The measured position of the $K^0_S$  mass peak allows a reliable evaluation of the possible bias in the $K^0_S$ four-momentum measurement.
 We selected  inclusive $K^0_S$ candidates satisfying the same $K^0_S$ selection criteria as in the $B^0 \to\psi^{(\prime)} \,K^0_S$  analysis; the momenta of   the selected inclusive $K^0_S$ candidates  were further restricted to be  from 1.55  to 1.85~GeV/$c$, which corresponds to the momentum range  of the $K^0_S$ mesons from $B^0 \to J/\psi\, K^0_S$ decays. Using this sample, we measured the mean reconstructed $K^0$
mass to be within 10~keV/$c^2$ of  the world average value of $497.672\pm0.031$~MeV/$c^2$~\cite{PDG}. However, we also observed a $\pm40$~keV/$c^2$ variation of the measured mean  $K^0$ mass depending on the radial position of the $K^0_S$ decay vertex. To assign the  systematic uncertainty, we conservatively took the $K^0$ mass shift to be 40~keV/$c^2$ and   added in quadrature  the 30~keV/$c^2$ uncertainty in the world average $K^0$ mass 
to obtain a  total  shift of 50~keV/$c^2$.
This  50~keV/$c^2$ variation  in the measured $K^0_S$ mass could be obtained   by varying   each daughter pion's momentum by  0.018\%; the resulting variation of the measured $B^0$ mass was 0.26~MeV/$c^2$, which we assigned as a systematic uncertainty due to the 
$K^0_S$ four-momentum measurement. This uncertainty in $M(B^0)$ has a contribution from the 
uncertainty  in  the world  average value of the $K^0$ mass, which partially limited the precision  of our  $K^0_S$  mass peak position measurement. 
In addition, we varied by one standard deviation the world  average $K^0$ mass value used 
for  the $K^0_S$ mass-constraint fit; the resulting    0.04~MeV/$c^2$ variation of the measured $B^0$ mass was added to the systematic uncertainty.

\stepcounter{syst_counter} 
(\roman{syst_counter})  {\it $K^+$ momentum measurement.} ---
Comparing the momentum spectra of the muons from inclusive $J/\psi$ decays 
and  the kaons from  $B^+ \to \psi^{(\prime)}\, K^+$ decays, we 
concluded that $J/\psi \to \mu^+ \mu^-$ decays provide excellent  calibration 
sample  for the study of  the  systematic uncertainty 
associated with the $K^+$ momentum measurement. 
As discussed above, the observed $+0.5\pm0.2$~MeV/$c^2$ shift of the  $J/\psi \to \mu^+ \mu^-$ mass peak corresponds to a systematic overestimation of the  muon momenta by 0.02\%. We decreased  the measured $K^+$ momenta by 0.02\%, which resulted in a $-0.32$~MeV/$c^2$ shift of the measured $B^+$ mass. We applied   a $-0.32$~MeV/$c^2$ correction to our final result and  assigned  $100\%$ of the correction value as the  systematic uncertainty. The ionization energy loss for muons from  inclusive $J/\psi$'s differs slightly from that for kaons from $B^+ \to \psi^{(\prime)}\, K^+$ decays. To account for the systematic  uncertainty due to this difference, we measured the $B^+$ mass using the pion Billoir fit  hypothesis for kaon tracks. The resulting  shift ($0.08$~MeV/$c^2$) was added in quadrature to the systematic 
uncertainty. 
Because of acceptance of the muon chambers, the muons pointing to the end-cap region of the detector are under-represented  in comparison with the  kaons from the $B^+ \to \psi^{(\prime)}\, K^+$ decays. The tracks with low transverse momentum are more likely to be affected by the  magnetic field inhomogeneity, thus providing an  additional source of systematic bias, which will not be taken into account  by  studying $J/\psi \to \mu^+ \mu^-$ decays.  
However,  if a $K^+$ track has low transverse momentum, then its track parameters  are poorly measured, and  the mass fit  naturally assigns a low weight to this $B^+$ candidate. 
We studied the possible systematic bias both by 
varying the measured $K^+$ momentum by 0.1\% for the $K^+$ tracks  with 
$|\cos\theta|>0.8$, where $\theta$ is the  angle between a track and the beam direction,  and by excluding these low angle  tracks altogether.   
The largest shift ($0.08$~MeV/$c^2$) was added in quadrature to the systematic 
uncertainty.

\stepcounter{syst_counter} 
(\roman{syst_counter})  {\it  Detector misalignment.} --- 
 The detector misalignment effects were studied with high-momentum muon tracks from   $e^+ e^- \to \mu^+ \mu^-$ events.  
We measured the mean of the transverse momentum difference between the  $\mu^+$ and  $\mu^-$ tracks. We also studied  the dependence  of the sum of the  $\mu^+$ and  $\mu^-$ momenta on azimuthal angle $\phi$ and polar angle $\theta$ of the $\mu^+$ track. We parametrized our findings in terms of an average as well as  $\phi$- and $\theta$-dependent false curvature. 
  We varied the measured curvature of the signal candidate tracks according to these parametrizations and found the detector misalignment effects to be  negligible for our $B$ mass measurements.

In conclusion, 
we have determined the  masses of  neutral and charged $B$ 
mesons with  significantly better precision than  any  previously published 
result~\cite{PDG}.
 We obtained  $M(B^0)=5279.1 \pm 0.7[{\rm stat}] \pm 0.3[{\rm syst}]$~MeV/$c^2$ and $M(B^+)=5279.1 \pm 0.4[{\rm stat}]  \pm 0.4[{\rm syst}]$~MeV/$c^2$. 
The systematic uncertainties for the $M(B^0)$ and  $M(B^+)$ measurements are  independent except for the small common  uncertainties
due to the imperfect knowledge of  the  $J/\psi$  and  $\psi(2S)$ masses (item (iv) in Table~I). Combining  our  $M(B^0)$ and $M(B^+)$ measurements with  the world average value of the mass difference $M(B^0)-M(B^+)=0.34\pm0.32$~MeV/$c^2$~\cite{PDG},  we obtained  $M(B^0)= 5279.2 \pm 0.5$~MeV/$c^2$ and  $M(B^+)= 5278.9 \pm 0.5$~MeV/$c^2$.  Although these $M(B^0)$ and $M(B^+)$ values are
more precise than the results given above, obviously they are 
 strongly correlated: the correlation coefficient
 is $\rho(M(B^0),M(B^+))=0.81$.

We gratefully acknowledge the effort of the CESR staff in providing us with
excellent luminosity and running conditions.
This work was supported by 
the National Science Foundation,
the U.S. Department of Energy,
the Research Corporation,
the Natural Sciences and Engineering Research Council of Canada, 
the A.P. Sloan Foundation, 
the Swiss National Science Foundation, 
and the Alexander von Humboldt Stiftung.

\

\
\
\


\begin{thebibliography}{99}



\bibitem{B-mass-ARGUS-CLEO}
H.~Albrecht {\it et al.}
(ARGUS Collaboration),
Z.\ Phys.\ C {\bf 48}, 543 (1990); 
M.S.~Alam {\it et al.}
(CLEO Collaboration),
Phys.\ Rev.\  D {\bf 50}, 43 (1994).


\bibitem{charge-conjugate}
Charge conjugate modes are implied; $\psi^{(\prime)}$ stands for  
$J/\psi$ and $\psi(2S)$.  


\bibitem{PDG}
C.~Caso {\it et al.} 
(Particle Data Group),
Eur.\ Phys.\ J.\ C {\bf 3}, 1 (1998).


\bibitem{Kubota:1992ww}
Y.~Kubota {\it et al.}
(CLEO Collaboration),
Nucl.\ Instrum.\ Meth.\ A {\bf 320}, 66 (1992).




\bibitem{Hill:1998ea}
T.S.~Hill,
Nucl.\ Instrum.\ Meth.\ A {\bf 418}, 32 (1998).
 
\bibitem{Billoir:1984mz}
P.~Billoir,
Nucl.\ Instrum.\ Meth.\ A {\bf 225}, 352 (1984).


\bibitem{GEANT} CERN Program Library Long Writeup W5013 (1993).




\end{thebibliography}
\end{document}